\documentclass[aps]{revtex4}

\usepackage{graphicx}
\usepackage{dcolumn}
\usepackage{bm}
\usepackage{color} 
\usepackage{amsmath, amssymb}
\begin{document}
\title{Single-Particle Spectral Density of a Bose Gas in the Two-Fluid 
Hydrodynamic Regime}

\author{Emiko Arahata}

\author{Tetsuro Nikuni}%

\affiliation{%
Department physics, Faculty of science, Tokyo University of Science, \\
1-3 Kagurazaka, Shinjuku-ku, Tokyo 162-8601, Japan}%
\author{Allan Griffin}
\affiliation{
Department of Physics, University of Toronto,
Toronto, Ontario, M5S 1A7, Canada}
\date{\today}
\begin{abstract}
In Bose supefluids, the single-particle Green's function can be directly related to the superfluid velocity-velocity correlation function in the hydrodynamic regime. An explicit expression for the single-particle spectral density was originally 
written down by Hohenberg and Martin in 1965, starting from the two-fluid equations for a superfluid. We give a simple derivation of their results. Using these results, we calculate the relative weights of first and second sound modes in the single-particle spectral density as a function of temperature in a uniform Bose gas. We show that the second sound mode makes a dominant contribution to the single-particle spectrum in relatively high temperature region. We also discuss the possibility of experimental observation of the second sound mode in a Bose gas by photoemission spectroscopy.
\end{abstract}
\maketitle 
\section{Introduction} 
In the low-frequency dynamics of superfluid Bose and Fermi atomic gases, 
the most dramatic effects related to superfluidity are described by Landau's two-fluid hydrodynamics analogous to the case of liquid $^4$He \cite{B_GNZ}. 
These equations only describe
the dynamics when collisions are sufficiently strong to produce a state of local thermodynamic equilibrium \cite{Hydro_Landau}. 
This requirement is usually summarized as $\omega \tau \ll 1$, where $\omega$ is the frequency of a collective mode and $\tau$
 is the appropriate relaxation rate. In this regime two sound modes
can be distinguished: the first sound mode consists of
an in-phase oscillation of the superfluid and normal fluid
components, while the second sound mode consists
of an out-of-phase oscillation of the superfluid and normal
fluid components. 
The occurrence of two distinct modes arises from the presence of both a superfluid component and normal
fluid component, which are coupled to each other. Recently, there has been renewed interest in second sound mode in superfluid Bose and Fermi gases \cite{T_Heiselberg_Unitarysound,T_C,E_Joseph_PRL98,E_second_Bose,E_second_BF}. The study of ultracold gases in collisional hydrodynamic regime has been
difficult because the density and the $s$-wave scattering
length are typically not large enough. 
However, Feshbach resonances allow ones to achieve conditions where the Landau two-fluid description is correct. 
Recent experiments have begun to observe sound propagation in trapped superfluid Fermi gases with a Feshbach resonance \cite{E_Kinast_PRL92,E_Bartenstein_PRL92,E_Joseph_PRL98}.  At unitarity, the magnitude of the $s$-wave scattering length that characterizes the interactions between fermions in different hyperfine states diverges ($|a_s|\to \infty$). Owing to the strong interaction close to unitarity, the dynamics of superfluid Fermi gases with a Feshbach resonance at finite temperatures are expected to be described by Landau's two-fluid hydrodynamic equations \cite{T_tau_Massignan,E_second_BF}. 

More recently, sound propagation in a Bose-condensed gas in a highly elongated (cigar-shaped) trap  has been observed in Ref.~\cite{E_second_Bose}, 
where the thermal cloud is in the hydrodynamic regime and thus the system is described by the two-fluid model.
In this experiment, the sound wave in a highly elongated trapped gas can be excited by a sudden modification of a trapping potential using the focused laser beam. The resulting
density perturbations propagate with a speed of sound. This experimental work has reported some success, with evidence for a second sound
mode in superfluid Bose gases, but first sound mode was not observed. In order to clearly demonstrate the two fluid dynamics of superfluids, 
it will be important to observe both first and second sound modes. 
In the case of a sound wave excited by a sudden modification of a trapping potential, 
the thermal density perturbations (first sound) is so small that one cannot distinguish small density perturbations from signal-to-noise in the thermal cloud in this regime \cite{E_second_Bose}. 

In general, the single-particle Green's function of Bose superfluids in the low-frequency, long-wavelength regime is directly
related to the superfluid velocity-velocity correlation function \cite{HM}, which is therefore coupled to first and second sound modes.
Making use of this exact relation, In this paper, we propose that first and second sound can be probed by measuring the single-particle spectral density.
This type of quantity is directory related to the tunneling current spectroscopy \cite{current_BEC}. Resent experience on ultracold Fermi gases by JILA group \cite{E_photo_Nature} showed that the momentum-resolved photoemission-type spectroscopy is a powerful technique to directly probe single-particle excitations of ultracold atomic gases. 

In Sec. II, we discuss the relation between superfluid velocity-velocity correlation function and the single-particle Green's function of Bose superfluids,
which was discussed by Hohenberg and Martin (which will be referred to as ``HM") ~\cite{HM}.  
We also give a simple derivation of the explicit expression for the single-particle Green's function in the two-fluid hydrodynamic regime, following the approach
analogous to the derivation of the density correlation function given in Ch.14 of Ref.~\cite{B_GNZ}. 
%


In Sec. III, we calculate the relative weights of first and second sound mode in the single-particle spectral density. 
For this purpose, we use the Hartree-Fock-Bogoliubov (HFB)-Popov approximation \cite{T_HFB-popov_Griffin}, for calculating various thermodynamic variables. 
We also compare the relative weights of first and second sound modes in both the 
single-particle spectral density and the dynamic structure factor

In Sec. IV, We show the single-particle spectral density in connection with the rf-tunneling current spectroscopy \cite{T_photo_Torma, T_photo_Tsuchiya, current_BEC}. We will show that both first and second sound mode can be observed by photoemission-type spectroscopy.
For comparison, we also show the he single-particle spectral density in the collisionless limit using the HFB-Popov approximation.


\section{Superfluid velocity-velocity correlation function and the
single-particle Green's function}
According to HM theory, the superfluid velocity-velocity correlation function in the non-dissipative hydrodynamic limit is given by \cite{HM}
 \begin{align}
    \chi_{v_s,v_s}({\bf q},\omega)=\frac{\left(\rho_s\lambda_{v_s,v_s}\omega^2
    -u_1^2u_2^2q^2 \right)q^2}{\rho_s(\omega^2-u_1^2q^2)(\omega^2-u_2^2q^2)} ,\label{KM_4.30}
            \end{align}
 where $u_1$ and $u_2$ are first and second sound velocities,
 which satisfy $u_1^2u_2^2=\frac{T\rho_s \bar s^2}{\rho_nc_v}\frac{\partial P}{\partial \rho}{\bigg |}_T$. Here $\rho_s$ and $\rho_n$ are superfluid and normal fluid densities and $\bar s$ is the entropy per unit mass.
 The explicit expression for $\lambda_{v_s,v_s}$ is given by
 \begin{align}
 \lambda_{v_s,v_s}= \frac{1}{\rho}\left[    
\left(\frac{\partial P}{\partial \rho}\right)_{\bar s}
-2\frac{T}{c_v}\frac{\bar s}{\rho}\left(\frac{\partial P}{\partial T}\right)_\rho+\frac{T\bar s^2}{c_v}\right]
\label{KM_4.30lambda}
     \end{align}
 In the low-frequency region, the single-particle Green's function $G({\bf q},\omega)$ is related to
the superfluid velocity-velocity correlation function through \cite{HM}
        \begin{align}
     G({\bf q},\omega)= \frac{n_0 m^2}{q^2} \chi_{v_s,v_s}({\bf q},\omega),
  \end{align}
  where $n_0$ is the condensate density.
 The single-particle spectral density $A({\bf q},\omega)$ 
 is related to the Green's function $G({\bf q},\omega)$ through
     \begin{align}
     G({\bf q},\omega)=\int_{-\infty}^{\infty} \frac{d\omega}{2\pi} \frac{A({\bf q}.\omega)}{\omega^\prime -\omega}
\label{G_A}
     \end{align} 
This gives
     \begin{align}
     A({\bf q},\omega)&=\frac{2\pi n_0m^2}{\rho_s}{\rm sgn}(\omega) 
\left[\frac{u_1^2\rho_s\lambda_{v_s,v_s}-u_1^2u_2^2}
{u_1^2-u_2^2}\delta(\omega^2-u_1^2q^2)+
\frac{u_2^2\rho_s\lambda_{v_s,v_s}-u_1^2u_2^2}{u_1^2-u_2^2}\delta(\omega^2-u_2^2q^2)
\right] \notag
     \\&
=
\frac{\pi n_0m^2}{\rho_sq}\Bigg\{
X_1
\left[\delta(\omega-u_1q)-\delta(\omega+u_1q)\right]+
X_2\left[\delta(\omega-u_2q)-\delta(\omega+u_2q)\right]\Bigg\},
     \label{eq_Aqw_1}
      \end{align}
where  $X_1$ and $X_2$ are defined by
\begin{align}
X_1 \equiv \frac{u_1(\rho_s\lambda_{v_s,v_s} -u_2^2)}{u_1^2-u_2^2}, ~~~
X_2 \equiv \frac{u_2(\rho_s\lambda_{v_s,v_s} -u_1^2)}{u_1^2-u_2^2}.
\label{eq_X1X2}
\end{align}

HM gives a systematic way to calculate various correlations
 functions in uniform superfluids in the two-fluid hydrodynamic regime.
However, their detailed derivation is quite involved, which closely follows the 
the earlier paper by Kadanoff and Martin \cite{KM} on a normal fluid.
In this section, we give an alternative derivation, analogous to 
Sec.14.3 of Ref.~\cite{B_GNZ} for the density-density correlation function.
We start with the non-dissipative two-fluid hydrodynamic equations:
\begin{equation}
\frac{\partial n}{\partial t} + \nabla\cdot{\bf j} = 0,
\label{eqfluid138}\end{equation}
\begin{align} 
m\frac{\partial {\bf j}}{\partial t} = 
-\nabla P,\label{eqfluid139}
\end{align}
\begin{align}
m\frac{\partial {\bf v}_s}{\partial t} = 
-\nabla\left(\mu+\frac{mv_s^2}{2}\right), \label{eqfluid142}\end{align}
\begin{align}\frac{\partial s}{\partial t} 
+\nabla\cdot(s{\bf v}_n)=0.
\label{eqfluid143}\end{align}
The total {\it mass} density and 
{\it mass} current are given by the sum of two components
\begin{align}
mn\equiv \rho = \rho_s+\rho_n, \label{eqfluid140}\end{align}
\begin{align} m{\bf j} \equiv \rho_s{\bf v}_s 
+\rho_n{\bf v}_n.
\label{eqfluid141}\end{align}

To calculate the superfluid velocity-velocity correlation function, we add a time-dependent
external current $\delta {\bf j}_{\rm ex}({\bf r},t)$ that is only coupled to the superfluid velocity ${\bf v}_s$.
That is, the continuity equation becomes
\begin{equation}
\frac{\partial n}{\partial t} + \nabla\cdot{\bf j} +\nabla\cdot \delta {\bf j}_{\rm ex}= 0,
\label{eqfluid138b}
\end{equation}
where as the entropy equation (\ref{eqfluid143}) is unchanged.

Taking time derivative of (\ref{eqfluid138b})Cwe obtain
\begin{equation}
\frac{\partial^2\rho}{\partial t^2}
=-m\nabla\cdot
\left(\frac{\partial {\bf j}}{\partial t}+\frac{\partial \delta {\bf j}_{\rm ex}}{\partial t}
\right)
=\nabla^2 P-m\nabla\cdot\frac{\partial \delta {\bf j}_{\rm ex}}{\partial t}
\label{eqrho2}
\end{equation}
Taking time derivative of (\ref{eqfluid143}) and linearize it in fluctuations, 
we obtain
\begin{equation}
\frac{\partial^2\delta s}{\partial t^2}=-s_0\nabla\cdot\frac{\partial{\bf v}_n}
{\partial t}
\end{equation}
One can show that
\begin{equation}
\rho_{n0}\frac{\partial \delta{\bf v}_n}{\partial t}
=
-\frac{n_{n0}}{n_0}\nabla\delta P-\frac{n_{s0}}{n_0}s_0\nabla\delta T
\end{equation}
and thus
\begin{equation}
\frac{\partial^2\delta s}{\partial t^2}=
\frac{s_0}{\rho_0}\nabla^2\delta P
+\frac{s_0^2}{\rho_0}\left(\frac{\rho_{s0}}{\rho_{n0}}\right)\nabla^2\delta T
\label{eqs2}
\end{equation}
This is the same as in the case without the external current, since the derivation does not involve the continuity equation.
It can be rewritten in terms of the local entropy per unit mass
$\bar s=s/\rho$. Using
\begin{equation}
\delta \bar s=-\frac{s_0}{\rho_0^2}\delta\rho+\frac{1}{\rho_0}\delta s,
\end{equation}
we see that
\begin{equation}
\frac{\partial^2 \delta \bar s}{\partial t^2}=
-\frac{s_0}{\rho_0^2}\frac{\partial^2\delta\rho}{\partial t^2}
+\frac{1}{\rho_0}\frac{\partial^2\delta s}{\partial t^2}
\label{eqbars}
\end{equation}
Using (\ref{eqrho2}) and (\ref{eqs2})
in (\ref{eqbars}), we obtain
\begin{eqnarray}
\frac{\partial^2 \delta \bar s}{\partial t^2}&=&
-\frac{s_0}{\rho_0^2}
\left(\nabla^2 \delta P-m\nabla\cdot\frac{\partial \delta {\bf j}_{\rm ex}}{\partial t}\right)
+\frac{1}{\rho_0}\left[\frac{s_0}{\rho_0}\nabla^2\delta P
+\frac{s_0^2}{\rho_0}
\left(\frac{\rho_{s0}}{\rho_{n0}}\right)\nabla^2\delta T\right] \nonumber \\
&=&\bar s_0^2
\left(\frac{\rho_{s0}}{\rho_{n0}}\right)\nabla^2\delta T
+\frac{\bar s_0}{n_0}
\nabla\cdot\frac{\partial \delta {\bf j}_{\rm ex}}{\partial t}
\end{eqnarray}

Let us now use $P$ and $T$ as independent variables.
We thus express fluctuations of $\rho$ and $s$ in terms of $P$ and $T$
as
\begin{equation}
\delta\rho=
\left(\frac{\partial\rho}{\partial P}\right)_{T}\delta P+
\left(\frac{\partial\rho}{\partial T}\right)_{P}\delta T
\end{equation}
\begin{equation}
\delta \bar s=
\left(\frac{\partial \bar s}{\partial P}\right)_{T}\delta P+
\left(\frac{\partial \bar s}{\partial T}\right)_{P}\delta T
\end{equation}
We then obtain
\begin{equation}
\left(\frac{\partial\rho}{\partial P}\right)_{T}\frac{\partial^2\delta P}{\partial t^2}
+\left(\frac{\partial\rho}{\partial T}\right)_{P}\frac{\partial^2\delta T}{\partial t^2}
-\nabla^2 P=-m\nabla\cdot\frac{\partial\delta{\bf j}_{\rm ex}}{\partial t}
\label{eq1}
\end{equation}
\begin{equation}
\left(\frac{\partial \bar s}{\partial P}\right)_{T}
\frac{\partial^2\delta P}{\partial t^2}+
\left(\frac{\partial \bar s}{\partial T}\right)_{P}\frac{\partial^2\delta T}{\partial t^2}
-\bar s_0^2\left(\frac{\rho_{s0}}{\rho_{n0}}\right)\nabla^2\delta T
=\frac{\bar s_0}{n_0}\nabla\cdot\frac{\partial\delta{\bf j}_{\rm ex}}{\partial t}
\label{eq2}
\end{equation}

We  consider an external current that excites modes of frequency $\omega$ and wavevector
${\bf q}$, namely
\begin{equation}
\delta {\bf j}_{\rm ex}({\bf r},t)=\delta {\bf j}_{\rm ex}({\bf q},\omega)e^{i({\bf q}\cdot{\bf r}-\omega t)},\label{Vex5}
\end{equation}
and look at the plane-wave solutions of Eqs.(\ref{eq1}),(\ref{eq2})
\begin{equation}
\delta P({\bf r},t)=\delta P_{{\bf q},\omega}e^{i({\bf q}\cdot{\bf r}-\omega t)},~~
\delta T({\bf r},t)=\delta T_{{\bf q},\omega}e^{i({\bf q}\cdot{\bf r}-\omega t)}.\label{Vex6}
\end{equation}
This gives two coupled algebraic equations
\begin{equation}
\left(
\begin{array}{c}
\delta P_{{\bf q},\omega} \\[8mm]
\delta T_{{\bf q},\omega}
\end{array}
\right)
\left(
\begin{array}{cc}
\displaystyle
\left(\frac{\partial\rho}{\partial P}\right)_{T}\omega^2-q^2
&
\displaystyle
\left(\frac{\partial\rho}{\partial T}\right)_{P}\omega^2 
\\[5mm]
\displaystyle
\left(\frac{\partial \bar s}{\partial P}\right)_{T}\omega^2
&
\displaystyle
\left(\frac{\partial \bar s}{\partial T}\right)_{P}\omega^2
-\bar s_0^2\left(\frac{\rho_{s0}}{\rho_{n0}}\right)q^2
\end{array}
\right)=
\left(
\begin{array}{c}
1
 \\[8mm]
 \displaystyle
-\frac{\bar s_0}{\rho_0}
\end{array}
\right)m\omega {\bf q}\cdot \delta {\bf j}_{\rm ex}({\bf q},\omega)
\end{equation}
We note that the determinant of the coefficient matrix is given by
\begin{eqnarray}
&&D(q,\omega) 
=\left[
\left(\frac{\partial T}{\partial \bar s}\right)_{\rho}
\left(\frac{\partial P}{\partial \rho}\right)_T
\right]^{-1} \nonumber \\
&&\times
\left\{\omega^4
-\left[\left(\frac{\partial P}{\partial \rho}\right)_{\bar s}
+\left(\frac{\rho_{s0}}{\rho_{n0}}\right)\bar s_0^2
\left(\frac{\partial T}{\partial \bar s}\right)_{\rho}
\right]\omega^2+\left(\frac{\rho_{s0}}{\rho_{n0}}\right)
\bar s_0^2
\left(\frac{\partial T}{\partial \bar s}\right)_{\rho}
\left(\frac{\partial P}{\partial \rho}\right)_T
\right\}.
\end{eqnarray}
In obtaining the final expression, we have made used of
the following relations:
\begin{equation}
\left[\frac{\partial(\rho,\bar s)}{\partial(P,T)}\right]^{-1}
=\frac{\partial(P,T)}{\partial(\rho,\bar s)}
=\frac{\partial(P,T)}{\partial(\rho,\bar s)}\frac{\partial(\rho,T)}{\partial(\rho,T)}
=\frac{\partial(P,T)}{\partial(\rho,T)}\frac{\partial(\rho,T)}{\partial(\rho,\bar s)}
=\left(\frac{\partial T}{\partial \bar s}\right)_{\rho}
\left(\frac{\partial P}{\partial \rho}\right)_T,
\label{eq:relation1}
\end{equation}
\begin{equation}
\left[\frac{\partial(\rho,\bar s)}{\partial(P,T)}\right]^{-1}
\left(\frac{\partial\bar s}{\partial T}\right)_P
=\frac{\partial(P,T)}{\partial(\rho,\bar s)}
\frac{\partial(P,\bar s)}{\partial (P,T)}
=\frac{\partial(P,\bar s)}{\partial(\rho,\bar s)}=
\left(\frac{\partial P}{\partial \rho}\right)_{\bar s},
\label{eq:relation2}
\end{equation}
\begin{equation}
\left[\frac{\partial(\rho,\bar s)}{\partial(P,T)}\right]^{-1}
\left(\frac{\partial  \rho}{\partial P}\right)_T
=\frac{\partial(P,T)}{\partial(\rho,\bar s)}
\frac{\partial(\rho,T)}{\partial (P,T)}
=\frac{\partial(\rho,T)}{\partial(\rho,\bar s)}=
\left(\frac{\partial T}{\partial \bar s}\right)_{\rho}.
\label{eq:relation3}
\end{equation}
One can also write the determinant in the compact form
\begin{equation}
D(q,\omega)=\frac{\partial(\rho,\bar s)}{\partial (P,T)}
(\omega^2-u_1^2q^2)(\omega^2-u_2^2q^2)
\end{equation}

The solution of the matrix equation is 
\begin{eqnarray}
\left(
\begin{array}{c}
\delta P_{{\bf q},\omega} \\[8mm]
\delta T_{{\bf q},\omega}
\end{array}
\right)
&=&\frac{1}{(\omega^2-u_1^2q^2)(\omega^2-u_2^2q^2)} 
\frac{\partial(P,T)}{\partial(\rho,\bar s)}
\nonumber \\
&&\times
\left(
\begin{array}{c}
\displaystyle
\left[\left(\frac{\partial \bar s}{\partial T}\right)_{P}
-\frac{\bar s_0}{\rho_0}\left(\frac{\partial\rho}{\partial T}\right)_{P}
\right]\omega^2 
-\bar s_0^2\left(\frac{\rho_{s0}}{\rho_{n0}}\right)q^2
\\[5mm]
\displaystyle
\left[-\frac{\bar s_0}{\rho_0}
\left(\frac{\partial\rho}{\partial P}\right)_{T}
-\left(\frac{\partial \bar s}{\partial P}\right)_{T}\right]
\omega^2
+\frac{\bar s_0}{\rho_0}q^2
\end{array}
\right)m\omega {\bf q}\cdot\delta {\bf j}_{\rm ex}({\bf q},\omega)
\label{eqTP}
\end{eqnarray}
We have thus obtained the expressions for the temperature
and pressure fluctuations.
Using the Gibbs-Duhem relation
\begin{equation}
n \delta\mu=\delta P-n \bar s \delta T,
\end{equation}
we can derive the expression for the fluctuation of the chemical
potential.
  \begin{eqnarray}
 && \rho \delta \mu_{{\bf q},\omega} (\omega^2-u_1^2q^2)(\omega^2-u_2^2q^2)
 \nonumber \\ 
&&=\left\{\left[\left(\frac{\partial P}{\partial \rho}\right)_{\bar s}-\frac{T}{c_v}\frac{\bar s}{\rho}\left(\frac{\partial P}{\partial T}\right)_\rho-\bar s \rho \left(\frac{\partial T}{\partial \rho}\right)_{\bar s}+\frac{T\bar s^2}{c_v} \right]\omega^2- \frac{T \bar s^2}{c_v} \frac{\rho}{\rho_{n0}}\left(\frac{\partial P}{\partial \rho}\right)_T q^2\right\}
\nonumber \\
&&~~~~~~~\times \omega {\bf q} \cdot \delta {\bf j}_{\rm ex}({\bf q},\omega)
 \label{eq:muj}
 \end{eqnarray}
In deriving the final expression,  we have used 
Eqs.~(\ref{eq:relation1})-(\ref{eq:relation3}) and analogous formulas.
The superfluid velocity can be then obtained from the linearized form of (\ref{eqfluid142}), i.e.
\begin{equation}
-im\omega \delta {\bf v}_{s{\bf q},\omega}=-i{\bf q}\delta\mu_{{\bf q},\omega}.
\label{muandvs}
\end{equation}
On the other hand, the superfluid velocity can be written in terms of the superfluid velocity-velocity correlation
function as
\begin{equation}
\delta {\bf v}_{s{\bf q},\omega}=
m\chi_{v_s,v_s}({\bf q},\omega)
\frac{{\bf q}}{q^2}{\bf q}\cdot\delta {\bf j}_{\rm ex}({\bf q},\omega).
\label{chivsvs}
\end{equation}
Using the relation (\ref{muandvs}),  one can write (\ref{chivsvs}) as
\begin{equation}
\delta \mu_{{\bf q},\omega}=
\chi_{v_s,v_s}({\bf q},\omega)
\frac{\omega}{q^2}{\bf q}\cdot\delta {\bf j}_{\rm ex}({\bf q},\omega).
\label{eq:muchi}
\end{equation}
Comparing (\ref{eq:muchi}) with (\ref{eq:muj}), we obtain the expression (\ref{KM_4.30}) for $\chi_{v_s,v_s}$, 
where $\lambda_{v_s,v_s}$ is now given by
\begin{equation}
\lambda_{v_s,v_s}=\frac{1}{\rho}\left[\left(\frac{\partial P}{\partial \rho}\right)_{s}-\frac{T}{c_v}\frac{\bar s}{\rho_0}
\left(\frac{\partial P}{\partial T}\right)_\rho-\bar s \rho \left(\frac{\partial T}{\partial \rho}\right)_{\bar s}+\frac{T\bar s^2}{c_v}\right].
\label{KM_4.30d}
\end{equation}
Finally, using the thermodynamic identity
\begin{equation}
\frac{T}{c_v}\frac{\bar s}{\rho}\left(\frac{\partial P}{\partial T}\right)_\rho=\bar s \rho\left(\frac{\partial T}{\partial \rho}\right)_{\bar s},
\label{identity}
\end{equation}
we can show that (\ref{KM_4.30d}) agrees with the HM expression (\ref{KM_4.30lambda}).
 \section{Amplitude of first and second sound}
 \subsection{HFB-Popov approximation}
\begin{figure}[htbp]
\centerline{\includegraphics[height=2.0in]{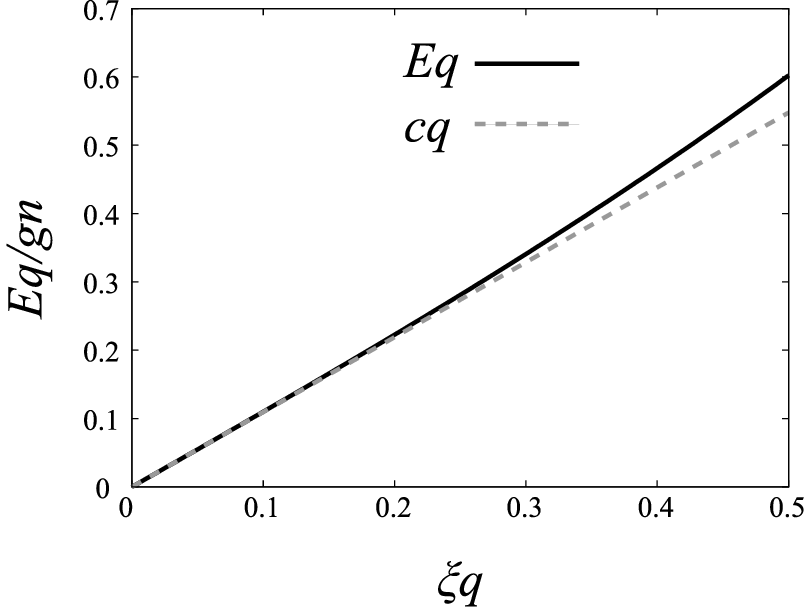}}
 \caption{Quasiparticle energy spectrum for $g^\prime=0.5$ at the temperature $T=0.5T_c^0$.The solid line is the quasiparticle excitation spectrum $Eq$ in Eq.~(\ref{E_q}) and the dashed line is the sound-like energy spectrum $E_q=c\hbar q$ with $c=\sqrt{gn_0/m}$}
  \label{Eq}
\end{figure}

 In order to calculate the single-particle Green's function explicitly, we
must specify a microscopic approximation for calculating thermodynamic
various variables.
Here we use the Hartree-Fock-Bogoliubov (HFB)-Popov approximation
\cite{T_HFB-popov_Griffin}.
For a uniform Bose gas, HFB-Popov approximation gives the quasiparticle excitation spectrum
  \begin{eqnarray}
  E_q=\sqrt{2gn_{\rm{0}}\epsilon_q^0+\left(\epsilon_q^0\right)^2}, \label{E_q}
  \end{eqnarray}
Here, $\epsilon_q^0=\frac{\hbar^2 q^2}{2m}$ is the single-particle energy of a noninteracting gas.
The noncondensate atom fraction is given by
\begin{eqnarray}
\tilde{N}=\sum_{q}\left[\frac{\epsilon_q^0+gn_0}{E_q}f\left(E_q\right)+\frac{1}{2}\left(\frac{\epsilon_q^0+gn_0}{E_q}-1\right)\right],~~N_0=N-\tilde{N}. \label{b1}
\end{eqnarray}
Here, $f(E)=1/(e^{\beta E}-1)$ is the Bose distribution function.
The quasiparticle amplitudes $u_q$ and $v_q$ are given by
 \begin{eqnarray} 
 u_q^2=\frac{1}{2}\left(\frac{\epsilon_q^0+gn_{\rm{0}}}{E_q}+1\right),\ \ \ \
 v_q^2=\frac{1}{2}\left(\frac{\epsilon_q^0+gn_{\rm{0}}}{E_q}-1\right). \label{u_v_0}
   \end{eqnarray}
Solving (\ref{E_q}) and (\ref{b1}) self-consistently, we obtain the condensate atom number $N_0$, 
noncondensate atom number $\tilde{N}$, and quasiparticle energy spectrum $E_q$.
The dimensionless interaction parameter is defined by
   \begin{eqnarray} 
 g^\prime=\frac{gn}{k_{\rm B}T_c^0},
  \end{eqnarray}
were $T_c^0$ is the BEC transition temperature of an ideal Bose gas.
In Fig.~\ref{Eq}, we plot the quasiparticle energy spectrum for the interaction $g^\prime=0.5$
at the temperature $T=0.5T_c^0$.
Here the wavenumber is normalized in terms of the healing length $\xi=\frac{\hbar}{\sqrt{2mgn}}$.
We will calculate the thermodynamic functions using the results obtained above.

In the HFB-Popov approximation, the thermodynamic potential is given by
\begin{eqnarray}
\Omega=-\mu n_0V +\frac{1}{2}gn_0^2V+k_{\rm{B}}T\sum_{q}\ln\left(1-e^{-\beta E_q}\right)+\sum_q E_q v_q^2.
\end{eqnarray}
The pressure is the given by $P=-\Omega/V$.
The entropy is given in terms of the quasiparticle excitation spectrum as
\begin{eqnarray}
S=k_{\rm{B}}\sum_q\Big\{\big[1+f(E_q)\big]\ln\big[1+f(E_q)\big]-f(E_q)\ln f(E_q)\Big\}.
\end{eqnarray}
Using these thermodynamic quantities, one can calculate the sound velocities.
The first and second sound velocities are given by
 \begin{eqnarray} 
 c^2_{12}=\frac{C_s^2+C_2^2}{2}\pm\sqrt{\left(\frac{C_s^2+C_2^2}{2}\right)^2-C_T^2C_2^2}, \label{u12}
\end{eqnarray}
where
 $C_s^2=\left(\frac{\partial P}{\partial \rho}\right)_{\bar{s}}$,  $C_T^2=\left(\frac{\partial P}{\partial \rho}\right)_{T}$, $C_2^2=\frac{\rho_{s0}}{\rho_{n0}}\frac{T\bar{s}^2_0}{\bar{C_v}}$. $C_s^2-C_T^2=\left(\frac{\partial s}{\partial \rho}\right)_T^2\frac{\rho^2T}{c_v}$. 

\begin{figure}[htbp]
\centerline{\includegraphics[height=2.5in]{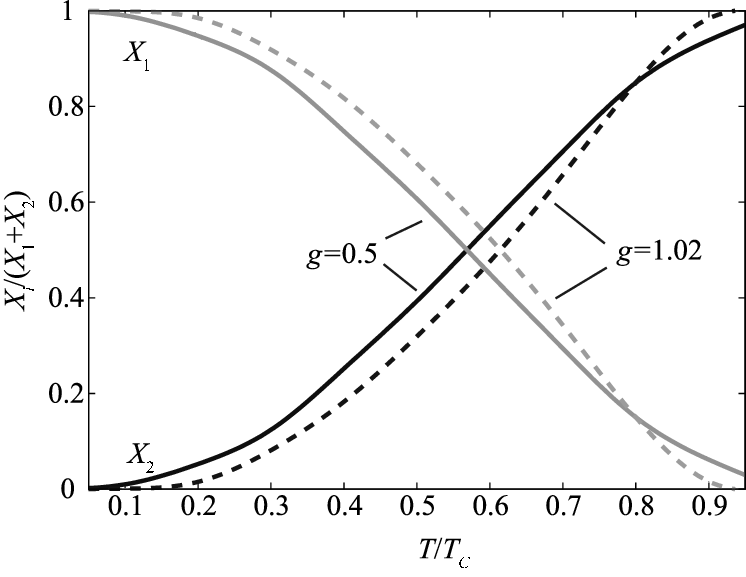}}
 \caption{Temperature dependence of relative weights of first and second sound modes $X_1$ and $X_2$ for $g^\prime=0.5$ and $g'=1.02$.}
  \label{chi_1b}
\end{figure}

Using the above results, we calculate the relative weights  of first and second sound in $A(\mathbf{q},\omega)$
 (see Eq.~\ref{eq_X1X2}). 
We take the interaction parameter from the experiment of Ref.~\cite{E_second_Bose},
which reports the observation of second sound.
In this experiment, total number of $^{23}$Na atoms $N=1.7\times 10^8$,
radial trap frequency $\omega_{\rm rad}/2\pi=95$Hz, and the aspect ratio
$\omega_{\rm rad}/\omega_{\rm ax}\approx 65$.
We estimate the interaction parameter for a uniform gas using the average density
of the trapped gas, and obtain $g'\sim 0.5~(na^3\sim 0.07)$.
In Appendix B, we evaluate the characteristic collisional relaxation time and confirm that one is well in the hydrodynamic regime at intermediate temperature. 
In Fig.~\ref{chi_1b}, we plot the the temperature dependence of the relative weights
for $g'=0.5$ and $g'=1.02$.
We see that there is no significant difference between the two results.
In both cases, the first sound mode is dominant at low
temperature, while the second sound is dominant at high temperature.
This can be understood as follows.
In the case of the superfluid velocity-velocity correlation, the external
perturbation is directly coupled to the condensate motion.
Therefore, the condensate mode should be always dominant in the
spectral weight in the case of weakly-interacting Bose gas.
At very low temperature (near $T=0$), the first sound mode is essentially
the condensate collective mode and the second sound mode is the
collective mode of quasiparticle excitations.
With increasing temperature, there is a crossover between two modes,
and the nature of the sound oscillations changes.
At high temperatures, the first sound mostly involves the noncondensate
oscillation, while the second sound mostly involves the condensate oscillation.
In Appendix A, we compare the result with the self-consistent Hartree-Fock(HF) approximation. we see that the qualitative behaviors are well captured by the HF approximation. Moreover, we can explicitly see that in a weakly-interacting Bose gas, the single-particle spectrum is dominated by the condensate mode.

\begin{figure}[htbp]
\centerline{\includegraphics[height=2.5in]{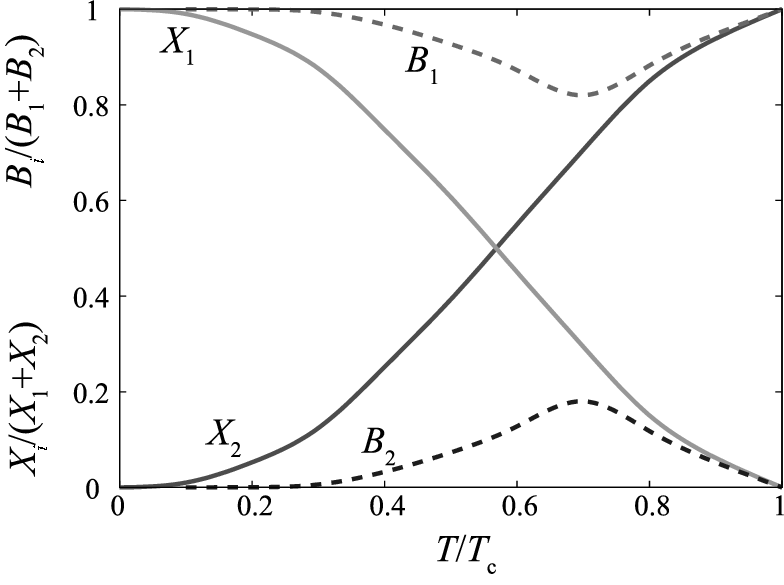}}
 \caption{Comparison of the relative weights of first and second
 sound modes between $A({\bf q},\omega)$ and $S({\bf q},\omega)$.}
  \label{wandz}
\end{figure}

In Fig.~\ref{wandz}, we compare the relative weights of first and second
sound modes in both the single-particle spectral density
$A({\bf q},\omega)$ and the dynamic structure factor $S({\bf q},\omega)$,
where
\begin{eqnarray}
S({\bf q},\omega)&=&[f(\omega)+1]\frac{q}{m}
\left\{\frac{Z_1}{u_1}\left[\delta(\omega-u_1q)+\delta(\omega+u_1q)\right]+\frac{Z_2}{u_2}\left[\delta(\omega-u_2q)+\delta(\omega+u_2q)\right]\right\} \nonumber \\
&=&\frac{q}{m}\left\{B_1(q)
\left[\delta(\omega-u_1q)+\delta(\omega+u_1q)\right]+
B_2(q)\left[\delta(\omega-u_2q)+\delta(\omega+u_2q)\right]\right\} ,
\end{eqnarray}
where
\begin{equation}
Z_1=\frac{u_1^2-v^2}{u_1-u^2},~~Z_2=1-Z_1,
\end{equation}
and
\begin{eqnarray}
B_1(q)=[f(u_1q)+1]\frac{Z_1}{u_1},~~  B_2(q)=[f(u_2q)+1]\frac{Z_2}{u_2} 
\end{eqnarray}
We see that in $S({\bf q},\omega)$, the first sound mode is dominant
at all temperatures. This is in sharp contrast with $A({\bf q},\omega)$,
where the second sound mode is dominant near $T_c$.

\section{Single-particle spectral density}
\begin{figure}[htbp]
\centerline{\includegraphics[height=3.0in]{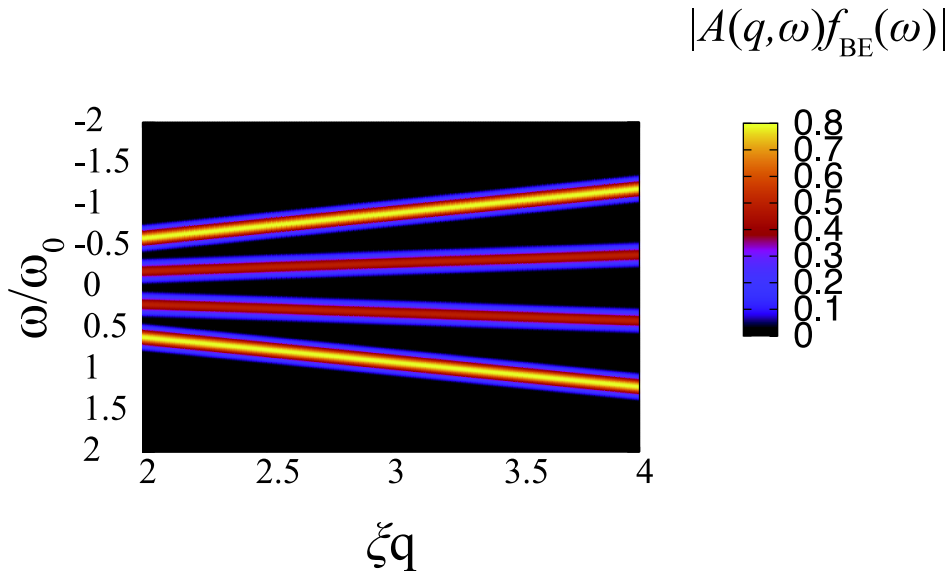}}
 \caption{ Plot of $A(\mathbf{q},\omega)$ as a function of $q$ and $\omega$
for $g^\prime=0.5~T=0.5T_c^0$. } 
  \label{Aqw}
\end{figure}

We now discuss the single-particle spectral density in connection with the experiment such as the 
the photoemission spectroscopy \cite{E_photo_Nature}.
In a uniform gas, the photoemission current is related to spectral weight $I_{\rm sw}(\mathbf{q},\omega) \propto A(\mathbf{q},\omega)f(\omega)$  \cite{T_photo_Torma,T_photo_Tsuchiya}.
In Fig.~\ref{Aqw}, we show $|A(\mathbf{q},\omega)f(\omega)|$ for $g'=0.5$ at 
the temperature $T/T_{c}^0=0.5$ as a function of
$q$ and $\omega$.
In the plot, the energy delta function is replaced by a gaussian with a finite width:
$\delta(\omega-c_1q)\to Ae^{a(\omega-c_1q)^2}$, where $a=1/4dq^2$
and $A=\sqrt{\frac{2\pi}{a}}$.
Some appropriate value of the width parameter $d$ is used.
In Fig.~\ref{Aqw}, the wavenumber $q$ and frequency $\omega$ are 
normalized in terms of the condensate healing lenght $\xi$ and the
mean-field frequency $\omega_0=gn/\hbar$.
For a parameter set we used, $\xi\approx 0.87\mu$m and $\omega_0\approx 2.7$kHz. The wavenumber and frequency are comparable to those in the recent experiment \cite{E_photo_Nature}.   

\begin{figure}[htbp]
\centerline{\includegraphics[height=2.0in]{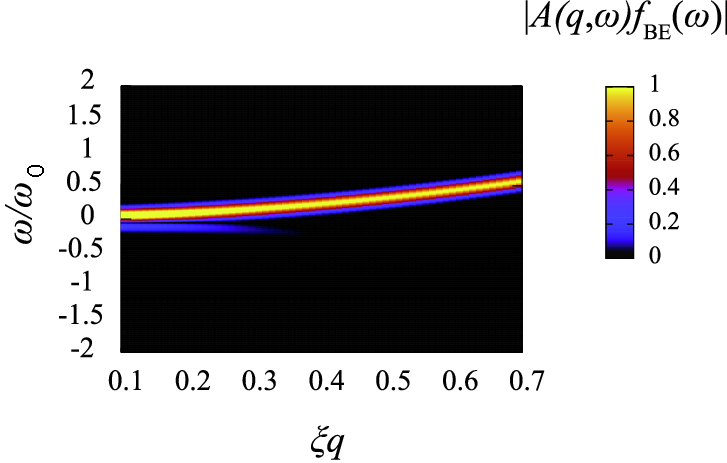}}
 \caption{ Plot of $A^{\rm Popov}(\mathbf{q},\omega)$ in the collisionless limit for
 $g^\prime=0.5,~T=0.5T_{\rm c}$.} 
\label{AqwPopov}
\end{figure}

For comparison, we also calculate $A(\mathbf{q},\omega)$ in the collisionless
limit using the HFB-Popov approximation:
 \begin{eqnarray}
 A^{\rm Popov}(\mathbf{q},\omega)=u^2_q\delta(\omega-E_q)-v^2_q\delta(\omega+E_q)
 \label{Aqw_Popov}
\end{eqnarray}
In Fig.~\ref{AqwPopov}, we plot this collisionless result for the same
coupling and temperature $g^\prime=0.5,~T=0.5T_{\rm c}^0$.
In the collisionless limit, there is only one sound mode.
Moreover, it is clear from the expression 
 (\ref{Aqw_Popov}) that the weights of $\omega =+E_q$ and
 $\omega=-E_q$ are different, since in general
 $u_q\neq v_q$.
 This is in contrast with the two-fluid hydrodynamic regime, where
 $\omega=+u_{1,2}q$ and $\omega=-u_{1,2}q$ have the same
 weights.
 In particular, for large $q$ one has $u_q \gg v_q$ and thus
 $A(\mathbf{q},\omega)$ only has the contribution from $\omega=+E_q$.
 In the opposite low-frequency limit, one has $u_p^2/v_p^2\to 1$.

\section{conclusion}
In this paper, we have discussed single-particle spectral densities of first and second sound in superfluid Bose gases. In order to obtain the thermodynamic quantities available for calculating the single-particle spectral densities of the first and second sound modes, we use the HFB-Popov approximation. 
We showed that that both first and second sound mode can be observed by photoemission spectroscopy near $T\sim0.5T_{\rm c}$. We showed that  the first sound mode is dominant at low temperature, while the second sound is dominant at high temperature.
With increasing temperature, there is a crossover between two modes,
and the nature of the sound oscillations changes. We hope that our results will stimulate further experiment by photoemission spectroscopy in a superfluid Bose gas in the two-fluid hydrodynamic regime.

We have compared the relative weights of first and second
sound modes in both the single-particle spectral density
$A({\bf q},\omega)$ and the dynamic structure factor $S({\bf q},\omega)$. We showed that in $S({\bf q},\omega)$, the first sound mode is dominant at all temperatures. This is in sharp contrast with $A({\bf q},\omega)$, where the second sound mode is dominant near $T_c$. 

For illustration, we also considered the self-consistent Hartree-Fock approximation. We showed that the qualitative behaviors are well captured by the HF approximation, 
but the quantitative details are quite different. 

Finally, we showed thespectral weight $|A(\mathbf{q},\omega)f(\omega)|$ as a function of $q$ and $\omega$. We found that both first and second sound mode can be observed by photoemission spectroscopy.

\section{ACKNOWLEDGMENTS}
We thank S. Tsuchiya for valuable comments. 
This research was supported by Academic Frontier Project (2005) of MEXT. E. A. is supported by a Grant-in-Aid from JSPS.
\appendix

\section{Self-consistent Hartree-Fock approximation and the
ZNG hydrodynamics}

For illustration, we consider the self-consistent Hartree-Fock approximation.
In this approximation, the
noncondensate density is given by
\begin{equation}
\tilde n=\frac{1}{\Lambda^3}g_{3/2}(z),
\label{HFeq1}
\end{equation}
where $\Lambda$ is the thermal de Broglie wavelength defined by
\begin{equation}
\Lambda=\left(\frac{2\pi\hbar^2}{mk_{\rm B}T}\right)^{1/2},
\end{equation}
and the fugacity $z$ is given by
\begin{equation}
z=\exp(-\beta g n_c),
\end{equation}
where $n_c=n-\tilde n$.
The pressure is given by
\begin{equation}
P=\tilde P+\frac{1}{2}g(n^2+2n \tilde n -\tilde n^2)
=\tilde P+\frac{1}{2}g(n_c^2+4\tilde n n_c + 2\tilde n^2),
\end{equation}
where $\tilde P$ is the kinetic pressure given by
\begin{equation}
\tilde P=\frac{k_{\rm B}T}{\Lambda^3}g_{5/2}(z)
\end{equation}
The entropy is given by
\begin{equation}
S=\frac{1}{T}\left(
\frac{5}{2}\tilde P+g\tilde n n_c\right)
\end{equation}
We can then calculate the various thermodynamic derivatives used in the
single-particle spectral function. 

Alternatively, one can use the linearized ZNG hydrodynamic equations \cite{B_GNZ,ZNG}, which are based on the self-consistent HF approximation,
to derive the superfluid velocity-velocity correlation function.
The coupled equations for the condensate and noncondensate 
variables are given by
\begin{eqnarray}
\frac{\partial\delta \tilde{n}}{\partial t} &=& -
\tilde{n}_0({\nabla}\cdot\delta{\bf v}_n)+\delta \Gamma_{12}, \label{greens_eq1}\\
m\tilde{n}_0\frac{\partial\delta{\bf v}_n}{\partial t} &=& -{\nabla}
\delta\tilde{P}-2g\tilde{n}_0{\nabla} (\delta\tilde{n}+\delta n_c), 
\label{greens_eq2}\\
\frac{\partial\delta\tilde{P}}{\partial t} &=& -\frac{5}{ 3}
\tilde{P}_0 ({\nabla}\cdot\delta{\bf v}_n)
\cdot{\nabla}\tilde{P}_0+\frac{2}{ 3} (\mu_{c0} - U_0) \delta
\Gamma_{12}\, ,
\label{greens_eq3}
\end{eqnarray}
\begin{eqnarray}
\frac{\partial\delta n_c}{\partial t} &=& - 
n_{c0}  ( {\nabla}\cdot \delta{\bf v}_c)
-\delta \Gamma_{12}, \label{greens_eq4}\\
m\frac{\partial\delta{\bf v}_c}{\partial t}
&=& -{\nabla}\delta\mu_c,
\label{greens_eq5}
\end{eqnarray}
\noindent
where
\begin{equation}
\delta \mu_c({\bf r}, t) =  g\delta n_c({\bf r}, t)
+ 2g\delta \tilde n({\bf r},t)\,.
\label{greens_eq6}
\end{equation}
The expression for $\delta \Gamma_{12}$ is given by
\begin{equation}
\delta \Gamma_{12}[\tilde f]=
-\frac{\beta_0 n_{c0}}{ \tau_{12}} \delta\mu_{\rm diff},~~~
\mu_{\rm diff} \equiv \tilde \mu - \mu_c.
\label{greens_eq7}
\end{equation}
In order to calculate the superfluid velocity-velocity correlation function, we
add a {\it time-dependent} external current $\delta {\bf j}_{\rm ex}({\bf r},t)$ to the
equation of motion for $\delta n_c$ given in (\ref{greens_eq4}).
That is, we use
\begin{equation}
\frac{\partial\delta n_c}{\partial t} = - 
n_{c0}  ( {\nabla}\cdot \delta{\bf v}_c)-\nabla\cdot \delta {\bf j}_{\rm ex}
-\delta \Gamma_{12}, \label{greens_eq8}\\
\end{equation}

To solve the hydrodynamic equations,
we introduce velocity potentials according to $\delta
{\bf v}_c \equiv {\nabla} \phi_c$ and $\delta {\bf v}_n \equiv {\nabla} \phi_n$.  In terms of these new variables, 
the equations for the condensate and the equations for the
noncondensate  can be combined to give
\begin{eqnarray}
m\frac{\partial^2 \phi_c }{ \partial t^2} &=& gn_{c0} \nabla^2 \phi_c +
2g\tilde n_0 \nabla^2 \phi_n +\frac{\sigma_H}{ \tau_\mu} \delta \mu_{\rm diff}
+g\nabla\cdot\delta {\bf j}_{\rm ex},
\label{greens_eq9} \\
m\frac{\partial^2 \phi_n }{ \partial t^2} &=& \left ( \frac{5\tilde P_0 }{
3\tilde n_0} + 2g\tilde n_0 \right ) \nabla^2 \phi_n +
2g n_{c0} \nabla^2 \phi_c - \frac{2\sigma_H }{ 3
\tau_\mu} \frac{n_{c0}}{\tilde n_0}
\delta \mu_{\rm diff}+2g\nabla\cdot \delta {\bf j}_{\rm ex}.
\label{greens_eq10}
\end{eqnarray}
Here $\delta\Gamma_{12}$ has been expressed in terms of  $\delta\mu_{\rm diff}$  using (\ref{greens_eq7}).
The equation of motion for $\delta\mu_{\rm diff}$ is given by (15.71)
\begin{equation}
\frac{\partial \delta\mu_{\rm diff}}{\partial t}=
\frac{2}{3}gn_{c0}\nabla^2\phi_n-
gn_{c0}\nabla^2\phi_c-g\nabla\cdot\delta {\bf j}_{\rm ex}
-\frac{\delta \mu_{\rm diff}}{\tau_{\mu}}.
\label{greens_eq11}
\end{equation}

We now consider an external current which excites modes of frequency
$\omega$ and wavevector ${\bf q}$
\begin{equation}
\delta {\bf j}_{\rm ex}({\bf r},t)=\delta {\bf j}_{\rm ex}({\bf q},\omega)e^{i({\bf q}\cdot{\bf r}
-\omega t)},
\end{equation}
and look for the plane-wave solutions
$\phi_{c,n}({\bf r},t) = \phi_{c,n,{\bf q},\omega} e^{i({\bf q}\cdot{\bf r} -\omega t)}.$
In this case, (\ref{greens_eq11}) reduces to
\begin{equation}
\delta \mu_{\rm diff} = \frac{\tau_\mu }{ 1-i\omega\tau_\mu}
\left[gn_{c0}\left ( \phi_c - \frac{2}{ 3}\phi_n \right ) q^2
-ig{\bf q}\cdot \delta {\bf j}_{\rm ex}\right].
\label{greens_eq13}
\end{equation}
Substituting this result into (\ref{greens_eq9}) and (\ref{greens_eq10}), we are left with two coupled equations
for the superfluid and normal fluid velocity potentials: 
\begin{eqnarray}
m\omega^2 \phi_{c,{\bf q},\omega} &=& gn_{c0}
\Bigg ( 1 - \frac{\sigma_H }{ 1-i\omega\tau_\mu} \Bigg ) q^2 \phi_{c,{\bf q},\omega}
+ 2g\tilde n_0 \left [ 1 + \frac{\sigma_H }{ 3
(1-i\omega\tau_\mu) } \frac{n_{c0} }{ \tilde n_0}
\right ] q^2 \phi_{n,{\bf q},\omega} \nonumber \\
&&-ig\Bigg ( 1 - \frac{\sigma_H }{ 1-i\omega\tau_\mu} \Bigg ) {\bf q}\cdot \delta {\bf j}_{\rm ex}({\bf q},\omega),
\label{greens_eq14}
\end{eqnarray}
and
\begin{eqnarray}
m\omega^2 \phi_{n,{\bf q},\omega} = \Bigg \{ \frac{5\tilde P_0 }{ 3\tilde n_0} &+& 2g\tilde
n_0 \left [ 1- \frac{2\sigma_H }{ 9 (1-i\omega\tau_\mu) } \frac{n_{c0}^2}{ \tilde n_0^2} \right ]\Bigg \} q^2 
\phi_{n,{\bf q},\omega} \nonumber \\ 
&+& 2gn_{c0} \left [ 1 + \frac{\sigma_H }{ 3 (1-i\omega\tau_\mu) } \frac{n_{c0} }{ \tilde n_0}
\right ] q^2 \phi_{c,{\bf q},\omega} \nonumber \\
&-& i2g \left [ 1 + \frac{\sigma_H }{ 3 (1-i\omega\tau_\mu) }\frac {n_{c0} }{ \tilde n_0}
\right ] {\bf q}\cdot \delta {\bf j}_{\rm ex}({\bf q},\omega).
\label{greens_eq15}
\end{eqnarray}

Taking the limit $\omega \tau_\mu \to 0$ of these coupled equations, we obtain
\begin{eqnarray}
m\omega^2 \phi_{c,{\bf q},\omega} &=& gn_{c0} ( 1 - \sigma_H ) q^2 \phi_{c,{\bf q},\omega}
+ 2g\tilde n_0 \left ( 1 + \frac{\sigma_H n_{c0}}{ 3 \tilde n_0} \right ) q^2 \phi_{n,{\bf q},\omega}\nonumber \\
&&-ig(1-\sigma_H){\bf q}\cdot\delta {\bf j}_{\rm ex}({\bf q},\omega),
 \label{greens_eq16} \\
m\omega^2 \phi_{n,{\bf q},\omega} &=& 
\Bigg [ \frac{5\tilde P_0}{ 3\tilde n_0} + 2g\tilde
n_0\left( 1 - \frac{2\sigma_H n_{c0}^2}{ 9\tilde n_0^2} \right) \Bigg ]
 q^2 \phi_{n,{\bf q},\omega}  \nonumber\\
&&{}{}{} + 2gn_{c0} \left ( 1 + \frac{\sigma_H n_{c0} }{ 3 \tilde n_0} \right ) q^2 \phi_{c,{\bf q},\omega}
-i2g\left ( 1 + \frac{\sigma_H n_{c0} }{ 3 \tilde n_0} \right ){\bf q}\cdot\delta {\bf j}_{\rm ex}({\bf q},\omega).
\label{greens_eq17}
\end{eqnarray}
It is useful to rewrite (\ref{greens_eq16}) and (\ref{greens_eq17}) in a
simple matrix form as
\begin{equation}
\left(
\begin{array}{cc} 
     \omega^2-v_2^2q^2 & -v_{21}^2q^2 \\
    -v_{12}^2q^2 & \omega^2-v_1^2q^2\\
      \end{array}\right)
\left(
   \begin{array}{c} 
     \phi_{c,{\bf q},\omega} \\
     \phi_{n,{\bf q},\omega} 
      \end{array}
  \right)
=-i\frac{{\bf q}\cdot \delta {\bf j}_{\rm ex}({\bf q},\omega)}{n_{c0}}
\left(
   \begin{array}{c} 
    v_{2}^2 \\ 
v_{12}^2 
      \end{array}
  \right),
\label{greens_eq18}
\end{equation}
where we have introduced new velocities
\begin{eqnarray}
v_2^2&=&\frac{gn_{c0}}{m}(1-\sigma_H),~~
v_{21}^2=\frac{2g\tilde n_0}{m}
\left(1+\frac{\sigma_Hn_{c0}}{3\tilde n_0}\right), \nonumber \\
v_{12}^2&=&\frac{2g n_{c0}}{m}
\left(1+\frac{\sigma_Hn_{c0}}{3\tilde n_0}\right),~~
v_1^2=\frac{5\tilde P_0}{3m\tilde n_0}+\frac{2g\tilde n_0}{m}
\left( 1-\frac{2\sigma_Hn_{c0}^2}{9\tilde n_0^2} \right).
\label{greens_eq19}
\end{eqnarray}
We note that these new velocities are related to the first and second sound velocities $u_1$ and $u_2$
through
\begin{equation}
u_1^2+u_2^2=v_1^2+v_2^2,~~u_1^2u_2^2=v_1^2v_2^2-v_{12}^2v_{21}^2.
\label{greens_eq20}
\end{equation}

Solving (\ref{greens_eq18}), we obtain
\begin{equation}
\left(
   \begin{array}{c} 
     \phi_{c,{\bf q},\omega} \\
     \phi_{n,{\bf q},\omega} 
      \end{array}
  \right)
=-i\frac{{\bf q}\cdot \delta {\bf j}_{\rm ex}({\bf q},\omega)}{n_{c0}}
\frac{1}{(\omega^2-u_1^2q^2)(\omega^2-u_2^2q^2)}
\left(
\begin{array}{cc} 
     \omega^2-v_1^2q^2 & v_{21}^2q^2 \\
    v_{12}^2q^2 & \omega^2-v_2^2q^2\\
      \end{array}\right)
\left(
   \begin{array}{c} 
    v_{2}^2 \\ 
v_{12}^2 
      \end{array}
  \right).
\label{greens_eq18b}
\end{equation}
We thus obtain
\begin{equation}
\phi_{c,{\bf q},\omega}=-i
\frac{{\bf q}\cdot\delta {\bf j}_{\rm ex}}{n_{c0}}
\frac{v_2^2\omega^2-(v_1^2v_2^2-v_{12}^2v_{21}^2)q^2}{(\omega^2-u_1^2q^2)(\omega^2-u_2^2q^2)}.
\label{greens_eq21}
\end{equation}
From (\ref{greens_eq21}), we obtain the superfluid velocity $\delta {\bf v}_c({\bf r},t)
=\delta {\bf v}_{c,{\bf q},\omega}e^{i({\bf q}\cdot{\bf r}-\omega t)}$, where
\begin{equation}
\delta {\bf v}_{c,{\bf q},\omega}=i{\bf q}\phi_{c,{\bf q},\omega}=
\frac{{\bf q}({\bf q}\cdot\delta {\bf j}_{\rm ex})}{n_{c0}}
\frac{v_2^2\omega^2-(v_1^2v_2^2-v_{12}^2v_{21}^2)q^2}{(\omega^2-u_1^2q^2)(\omega^2-u_2^2q^2)}.
\label{greens_eq22}
\end{equation}
This result can be written in terms of the superfluid velocity-velocity correlation function,
defined as
\begin{equation}
\delta v_{c\mu,{\bf q},\omega}=m\sum_{\nu}\chi_{v_sv_s}^{\mu\nu}({\bf q},\omega)\delta j_{{\bf q},\omega,\nu}.
\label{greens_eq23}
\end{equation}
Comparing (\ref{greens_eq23}) with (\ref{greens_eq22}), we find
\begin{equation}
  \chi_{v_sv_s}^{\mu\nu}({\bf q},\omega)=\frac{q_{\mu}q_{\nu}}{mn_{c0}}
\frac{v_2^2\omega^2-u_1^2u_2^2q^2}{(\omega^2-u_1^2q^2)(\omega^2-u_2^2q^2)}.
\label{greens_eq24}
\end{equation}
The longitudinal part of the superfluid velocity-velocity correlation function is
given by
\begin{equation}
  \chi_{v_sv_s}({\bf q},\omega)=\frac{q^2}{mn_{\rm c0}}
\frac{v_2^2\omega^2-u_1^2u_2^2q^2}{(\omega^2-u_1^2q^2)(\omega^2-u_2^2q^2)}.
\label{greens_eq25}
\end{equation}
Comparing this result with the general expression (\ref{KM_4.30}), 
we see that in the HF approximation one has $\rho_s\lambda_{v_sv_s}=v_2^2$ with $\rho_s=mn_{\rm c0}$.
The single-particle Green's function is given by
\begin{equation}
G({\bf q},\omega)=\frac{n_{\rm c0}m^2}{q^2}  \chi_{v_sv_s}({\bf q},\omega)=
\frac{m(v_2^2\omega^2-u_1^2u_2^2q^2)}{(\omega^2-u_1^2q^2)(\omega^2-u_2^2q^2)}.
\label{greens_eq26}
\end{equation}
This Green's function can be expressed in terms of the single-particle spectral density
through the relation (\ref{G_A}).
In this case, we have
\begin{eqnarray}
A({\bf q},\omega)&=&2\pi m~
{\rm sgn}(\omega)
\left[\frac{u_1^2(v_2^2-u_2^2)}{u_1^2-u_2^2}\delta(\omega^2-u_1^2q^2)
+\frac{u_2^2(u_1^2-v_2^2)}{u_1^2-u_2^2}\delta(\omega^2-u_2^2q^2)\right]\nonumber \\
&=&\frac{\pi m}{q}\Bigg\{X_1\left[\delta(\omega-u_1q)-\delta(\omega+u_1q)\right]
+X_2\left[\delta(\omega-u_2q)-\delta(\omega+u_2q)\right]\Bigg\},
\label{greens_eq30}
\end{eqnarray}
where
\begin{equation}
X_1=\frac{u_1(v_2^2-u_2^2)}{u_1^2-u_2^2},~~~
X_2=\frac{u_2(u_1^2-v_2^2)}{u_1^2-u_2^2}.
\end{equation}
The above expressions should be compared with the general expressions (\ref{eq_X1X2}).
If we use the expression for $u_2$ to order $g^2$ given by (15.86), one has
\begin{equation}
v_2^2=u_2^2+\Delta v^2,~~~\Delta v^2\equiv \frac{gn_{c0}}{m}\frac{3m\tilde n_0}{5\tilde P_0}\frac{4g\tilde n_0}{m},
\label{greens_eq31}
\end{equation}
and the weights $X_1$ and $X_2$ can be approximated as
\begin{equation}
X_1\approx \frac{u_1\Delta v^2}{u_1^2-u_2^2},~~
X_2\approx\left(1-\frac{\Delta v^2}{u_1^2-u_2^2}\right)u_2
\label{greens_eq32}
\end{equation}
We find from (\ref{greens_eq32}) that the contribution at the second sound mode $\omega^2=u_2^2q^2$ is
larger than the contribution at the first sound mode $\omega^2=u_1^2q^2$.

\begin{figure}[htbp]
  \centerline{\includegraphics[height=2in]{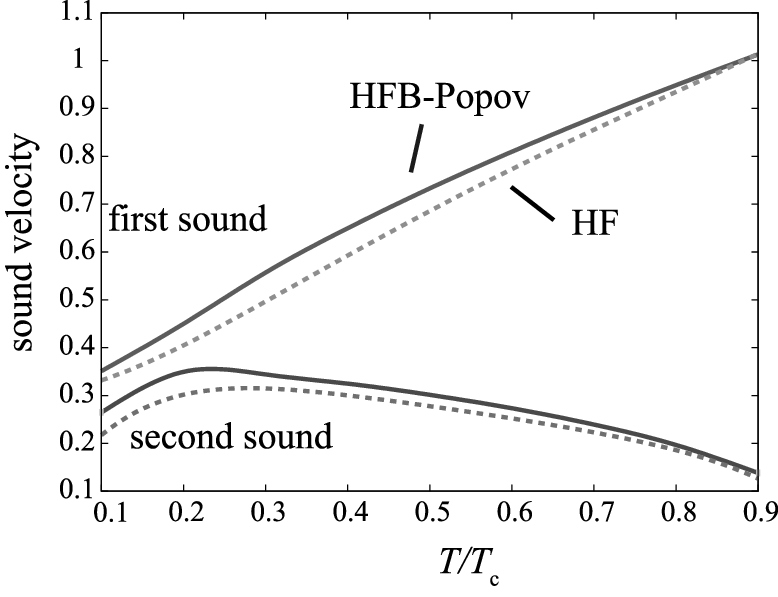}}
 \caption{Sound velocities of the first sound and second sound modes.
dash Lines show the results from the self-consistent HF approximation,
solid lines are the results from 
the HFB-Popov approximation.$g'=0.5$.}
\label{c1c2}
\end{figure}

\begin{figure}[htbp]
  \centerline{\includegraphics[height=2in]{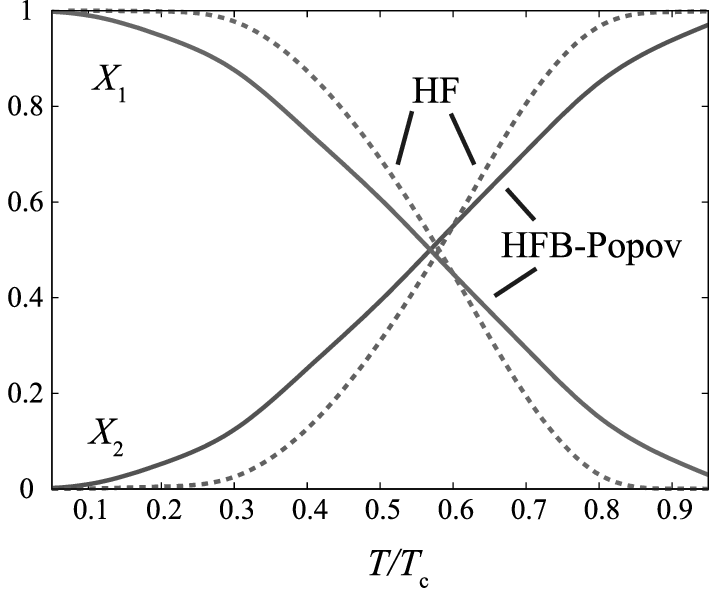}}
 \caption{The temperature dependence of the amplitudes $X_1,X_2$. 
 dash Lines show the results from the self-consistent HF approximation.
Solid lines are results from the HFB-Popov approximation.
 $g'=0.5$.}
 \label{chi2}
\end{figure}

Figure \ref{c1c2} compares the sound velocities calculated by the HF approximation
with those calculated by the HFB-Popov approximation.
Figure \ref{chi2} compares $X_1$ and $X_2$ calculated by the HF approximation
with those calculated by the HFB-Popov approximation.
We see that the qualitative behaviors are well captured by the HF approximation, but
the quantitative details are different.

\section{Validity of Hydrodynamics}
\begin{figure}[htbp]
\centerline{\includegraphics[height=2.0in]{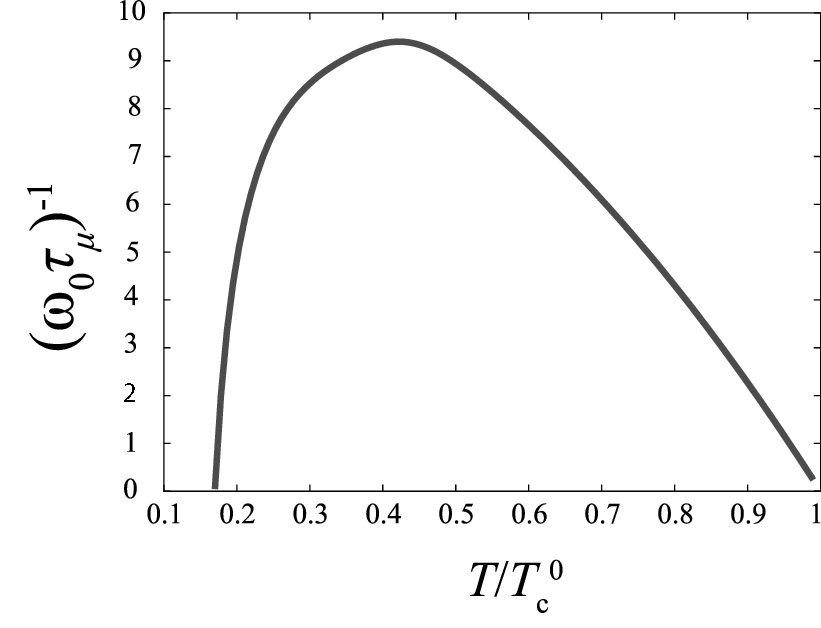}}
 \caption{ Temperature dependence of the relaxation time $\tau_{\mu}$.
 $g^\prime=0.5$.} 
\label{taumu}
\end{figure}

Here we discuss the validity of using Landau's two-fluid hydrodynamics in the
experiment of Ref.~\cite{E_second_Bose}.
In Fig.~\ref{taumu}, we plot the temperature dependence of the collisional relaxation time $\tau_{\mu}$ defined in Ref.~\cite{ZNG,T_tau_Nikuni}. This relaxation time $\tau_\mu$ describes the rate of equilibration of the condensate and noncondensate chemical potential. The condition for the hydrodynamic regime is described by $\omega\tau_{\mu}\le 1$ where $\omega$ is the frequency of the collective mode.
We see that the collective mode of the frequency $\omega\sim gn/\hbar$ is well
within the hydrodynamic regime in the intermediate temperature region $T\sim 0.5 T_{\rm c}$.

\bibliographystyle{apsrev}

\end{document}